\documentclass[12pt]{article}
\usepackage{amsmath,amssymb,exscale}  
\input epsf
\usepackage{axodraw}

\def\gappeq{\mathrel{\rlap {\raise.5ex\hbox{$>$}}
{\lower.5ex\hbox{$\sim$}}}}
\def\lappeq{\mathrel{\rlap{\raise.5ex\hbox{$<$}}
{\lower.5ex\hbox{$\sim$}}}}
   
\def\dsl{\hbox{/\kern-.6800em$D$}}
\def\rsl{\hbox{/\kern-.6800em$R$}}

\begin{document}
\pagestyle{empty}
\begin{flushright}
UAB-FT-538\\
UMN-TH-2126/03
\end{flushright}
\vspace*{5mm}

\begin{center}
{\Large\bf The Standard Model Partly Supersymmetric}\\
\vspace{1.0cm}

{{\sc Tony Gherghetta}$^a$ and {\sc Alex Pomarol}$^{b}$}\\
\vspace{.6cm}
{\it\small $^{a}$School of Physics and Astronomy,
University of Minnesota,\\
Minneapolis, MN 55455, USA}\\
{\it\small $^{b}$IFAE, Universitat Aut{\`o}noma de Barcelona,\\
08193 Bellaterra (Barcelona), Spain}
\vspace{.4cm}
\end{center}

\vspace{1cm}
\begin{abstract}
We present a novel class of theories where supersymmetry is only preserved
in a partial (non-isolated) sector. The supersymmetric sector consists
of CFT bound-states that can coexist with fundamental states  
which do not respect  supersymmetry. 
These theories arise from the 4D holographic interpretation  of
5D theories in a slice of AdS where supersymmetry is broken on the UV boundary.
In particular, we consider the Standard Model where only the Higgs sector 
(and possibly the top quark) is supersymmetric. The Higgs mass-parameter 
is then protected by supersymmetry, and consequently the electroweak 
scale is naturally smaller than the composite Higgs scale. 
This not only provides a solution to the hierarchy problem, but  
predicts a ``little'' hierarchy between the electroweak and new physics scale.
Remarkably, the model only contains a single supersymmetric partner, 
the Higgsino (and possibly the stop), and as in the usual MSSM, predicts 
a light Higgs boson.
\end{abstract}
\vfill
\begin{flushleft}
\end{flushleft}
\eject
\pagestyle{empty}
\setcounter{page}{1}
\setcounter{footnote}{0}
\pagestyle{plain}


\section{Introduction}

In quantum field theories it is unnatural to impose symmetries that 
are only restricted to certain parts of the Lagrangian.
This is because at the quantum level, interactions with sectors that are 
not symmetric will, in general, spoil the underlying symmetry of the 
symmetric sectors. In fact these quantum effects are generally 
divergent, which signals that one must include, from the beginning, 
all possible non-symmetric terms in the Lagrangian to absorb the 
infinities. Thus, in general, 
symmetries can only be maintained if the full theory 
respects them.

It is for this reason that in supersymmetric theories, which
are advocated to solve the hierarchy problem, supersymmetry  
must be respected in all sectors of the theory. If there is a sector 
where supersymmetry is not realized below the scale $\Lambda$, and is 
coupled to another supersymmetric sector with coupling $g$, then a 
fermion-boson mass splitting of order $\frac{g}{4\pi}\Lambda$ 
will be induced in the supersymmetric sector. Therefore to maintain 
supersymmetry in the presence of a nonsupersymmetric sector, either 
$g$ must be very tiny (decoupling limit), or $\Lambda$ is small 
(supersymmetry is preserved in the full theory down to low energies).

In this work we will present theories which are an exception to the 
above general argument. These theories will consist of two sectors:
a (super)symmetric sector that contains bound-states 
of a  spontaneously broken conformal field theory  (CFT),
and a  non-(super)symmetric sector which contains fundamental states.
Using the AdS/CFT correspondence, we will show
that the CFT bound-states decouple from the non-supersymmetric sector
at energies above  $1/L$,  where $L$ is the size of the bound-states 
(the scale of the conformal symmetry breaking). 
Thus, the bound states are insensitive to (super)symmetry breaking effects
at high energies.

This scenario allows us to have a non-supersymmetric sector 
coexisting with the supersymmetric sector, even though the coupling $g$
is of order one, and the scale $\Lambda$ is large.
As an interesting application~\cite{luty}, 
we will consider the Standard Model (SM) 
where only the Higgs sector is (approximately) supersymmetric.
This enables one to have a prediction for the tree-level 
Higgs potential. For example, the quartic coupling is determined by 
supersymmetry to be ${(g^2+g^{\prime 2})}/{8}$, whereas in the 
minimal supersymmetric SM (MSSM), this leads to a light Higgs boson 
mass. Furthermore, the Higgs 
mass-parameter, that determines the electroweak scale, is protected 
by supersymmetry down to low-energies
$\lappeq 1/L$. 
This mass
 can  be induced at
the quantum level by SM fields,  giving
\begin{equation}
      m_{EW}\sim\frac{g}{4\pi}\frac{1}{L}\ll \Lambda\sim M_P\, .
\end{equation}
Thus we see that these theories provide a rationale for having the   
electroweak scale naturally smaller than the composite Higgs scale $1/L$, 
which in turn can be much smaller than the Planck scale.
In fact, this ``little hierarchy'' between the electroweak scale and the scale
of new physics (which in our scenario is $1/L$) is currently suggested by 
present collider experiments. 
This has also motivated other theoretical 
models with this property, such as TeV extra dimension models or 
the ``little Higgs'' models \cite{little}.

\section{Partly (super)symmetric theories}

To understand why (super)symmetric CFT bound-states are not sensitive
to supersymmetry breaking effects at high energy scales, we will use
the AdS/CFT correspondence, and consider the 5D anti de-Sitter (AdS$_5$)
 point of view.
In this 5D dual picture the possibility of having partly (super)symmetric 
theories is very simple to understand.

\begin{figure}[t]
        \begin{center}
        \begin{picture}(230,190)(0,0)
            \Text(15,170)[l]{$y=0$}

            \Line(20,0)(20,150) 
             \Line(70,20)(70,170)
             \Line(20,0)(70,20)
             \Line(20,150)(70,170)

          \Text(100,30)[l]{\large AdS$_5$}

          \Text(25,50)[l]{No-Susy}
          \Text(25,40)[l]{Sector}
          \Text(180,50)[l]{Susy}
          \Text(180,40)[l]{Sector}

            \Text(165,170)[l]{$y=\pi $}
              \Line(170,0)(170,150) 
             \Line(220,20)(220,170)
             \Line(170,0)(220,20)
             \Line(170,150)(220,170)

              \DashLine(190,95)(200,65){3}
                \DashLine(190,95)(200,130){3}

               \PhotonArc(114,30)(100,40,140){3}{7} 
               \PhotonArc(114,159)(100,220,320){3}{7}
                 \Vertex(38,93){2.5}
                \Vertex(190,95){2.5}
  \end{picture}
\caption{\it Bulk contributions to the supersymmetric
sector at $y=\pi$.}
\end{center}
\end{figure}
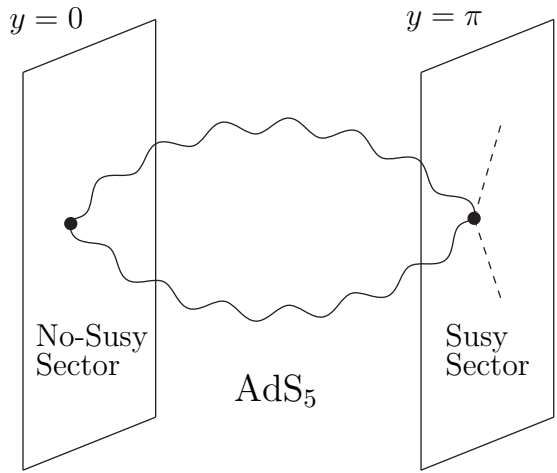

Let us then  start by considering a quantum field theory in
a slice of 5D AdS \cite{rs} (Fig.~1).
The 5D metric is given by
\begin{equation}
   ds^2 = a^2(y) dx^2 + R^2 dy^2\, ,
\label{metric}
\end{equation}
where $a(y)= e^{-kR y}$ is the warp factor,  $1/k$ is 
the AdS curvature radius, and $\pi R$ is the proper length of the extra 
dimension $y$. This 5D theory  has  two 4D boundaries at $y=0$ and $y=\pi$.
Due to the warp factor, energy scales on the $y=\pi$ boundary are reduced by 
a factor $a(\pi)$  compared to those on the $y=0$ boundary (this can 
be thought of as a redshift due to the warped space).
We will suppose that the full 5D theory is supersymmetric, and that
supersymmetry is broken only on the $y=0$ boundary at a scale of order 
the cutoff scale $\Lambda$ of the theory. 
Scalar fields localized on the $y=0$ boundary will receive masses of 
order $\Lambda$. Similarly, bulk scalar fields will receive 
boundary masses of this order. However, scalar fields living on the 
$y=\pi$ boundary, will only know about the breaking of supersymmetry 
by loop effects of bulk fields. Consequently, the contribution to their 
masses will arise from bulk fields propagating from the $y=\pi$ boundary to 
the $y=0$ boundary (where supersymmetry is broken) as shown in Fig.~1.
This is a non-local effect which gives a finite contribution to the 
scalar masses. What is  the order of magnitude of these corrections?
A rough estimate of the energy involved in the virtual  contribution
to the scalar mass is given by $E\sim 1/\tau\sim ka(\pi)$,
where $\tau$ is the time it takes to propagate from one boundary to the other
(the conformal distance). After multiplying by the coupling of the bulk 
field to the $y=0$ boundary, $g_0(E)$, and the  
$y=\pi$ boundary, $g_\pi(E)$, we obtain an estimate of the induced scalar mass
\begin{equation}
   m^2\sim \frac{g_0(\frac{1}{\tau})g_\pi(\frac{1}{\tau})}{16\pi^2} 
    a^2(\pi)k^2\, .
\label{estimate}
\end{equation}
Notice that due to the warp factor $a(\pi)$ this mass can be very small,
and consequently fields on the $y=\pi$ boundary
receive supersymmetry-breaking masses much smaller
than the cutoff scale, $\Lambda$. 
Of course, this is expected since we know that
energy scales on the $y=\pi$ boundary are
redshifted with respect to those on the $y=0$ boundary. 
More interestingly, it is the interpretation  of this effect 
in the  dual 4D theory to which  we now turn to.

The above theory has a 4D interpretation based on the 
AdS/CFT correspondence \cite{adscft}.
This correspondence relates the 5D AdS theory to a
strongly coupled 4D CFT with a large number of ``colors'' $N_c$.
The  5D bulk fields at the $y=0$  boundary,   $\Phi(x)$, 
are identified as sources of  CFT operators
\begin{equation}
{\cal L}=g\, \Phi(x) {\cal O}(x)\, ,
\label{sources}
\end{equation}
where the mass of $\Phi$ is related to the dimension of the
operator ${\cal O}$. The boundary at $y=0$  
corresponds to an ultraviolet (UV) cutoff at $p=k$  in the 4D CFT
\cite{gu}, while
the boundary at $y=\pi$ corresponds to an infrared (IR) cutoff at $p=k a(\pi)$
\cite{apr,rzp}.
Therefore, a slice of the bulk AdS space corresponds in 4D to a slice 
of CFT in momentum-space. Due to the term in Eq.~(\ref{sources}),  the
CFT can generate a kinetic term for the sources $\Phi(x)$ which become 
dynamical, and these must then be included in the theory as extra 
``fundamental'' fields.
The IR breaking of the CFT theory introduces a mass gap of 
order $1/L=k a(\pi)$, where bound states, $M$ (``mesons'' 
or ``baryons''), are formed. For  large $N_c$ \cite{largen}, 
we know that the infinite number of 
bound-states are weakly coupled, and  have a mass-spacing of order $ka(\pi)$.

The fundamental fields $\Phi$, and the CFT bound-states are not mass 
eigenstates, since, due to Eq.~(\ref{sources}), they will mix with each other.
However, after a rediagonalization, one can obtain the mass eigenstates. 
In this basis we can map these eigenstates into the zero-mode, and 
Kaluza-Klein (KK) modes of the dual 5D theory. The question of whether
the mass eigenstates are fundamental or CFT bound-states depends
on the amount of mixing or where the fields are localized. 
The Kaluza-Klein states always have wave-functions which are peaked towards 
the boundary at $y=\pi$. Thus, they will have a small wave-function
overlap with the fundamental fields $\Phi$, since by the AdS/CFT 
correspondence $\Phi$ is associated with the 4D field localized on the 
boundary at $y=0$. Thus, to a good approximation, the KK-states always 
correspond to the CFT bound-states. On the other hand, fields living  
completely on the boundary at $y=\pi$ will correspond  
to pure CFT bound-states because they cannot mix with
the fundamental fields $\Phi$.

\begin{figure}[t]
    \begin{minipage}[t]{0.5\linewidth}
        \begin{center}
        \begin{picture}(40,60)(0,0)
            \Photon(0,20)(40,20){3}{7}
            \Line(40,22)(80,22)
            \Line(40,18)(80,18)
           \Line(0,22)(-40,22)
            \Line(0,18)(-40,18)
            \Vertex(40,20){3}
            \Vertex(0,20){3}
            \Text(-45,22)[r]{$M$}
            \Text(84,22)[l]{$M$}
            \Text(24,27)[br]{$\Phi$}
        \end{picture}
        \caption{\it Tree-level
correction to the\hfill\break   mass of $M$.}
\label{massV}
        \end{center}
    \end{minipage}
     \begin{minipage}[t]{0.5\linewidth}
        \begin{center}
       \begin{picture}(20,60)(0,0)
            \Line(-30,22)(50,22)
            \Line(-30,18)(50,18)
                      \Text(-35,22)[r]{$M$}
            \Text(55,22)[l]{$M$}
           \Text(35,60)[br]{$\Phi$}
             \PhotonArc(10,42)(20,0,360){2}{20}
       \Vertex(10,20){8}
         \end{picture}
            \caption{\it  One-loop   
correction to the mass of $M$.}
\label{massS}
\end{center}
    \end{minipage}

\end{figure}

By supersymmetrizing the theory, the scalar bound-states 
can be massless since the scalar masses are then protected by supersymmetry. 
However, if in the 5D AdS theory we break supersymmetry at the $y=0$ 
boundary, then this corresponds in the 4D dual theory to breaking 
supersymmetry at the UV cutoff scale
in the  $\Phi$ sector.
The CFT  sector, however, is coupled 
to  $\Phi$ via Eq.~(\ref{sources}) and therefore it  will also feel
the breaking of supersymmetry.
Whenever $\langle 0|{\cal O}|M\rangle\not=0$, we have that
 $M$ and $\Phi$ mix.
The states $M$ will then receive a (tree-level) correction
to their masses that will not respect supersymmetry. For small
mixing this is giving by the diagram of Fig.~2. 
This mixing is completely negligible when $M$ is the CFT bound-state 
dual to a field living on the boundary at $y=\pi$.
In this case the dominant supersymmetry-breaking
effect will arise at the  one-loop level as shown in Fig.~3.
Nevertheless, we learnt from the AdS$_5$ dual theory 
that these contributions are generated at scales $\lappeq 1/L$.
Qualitatively, this can also be  understood in the CFT picture. 
The bound-state $M$ is a CFT lump of size $L$
that  decouples from  fields of wavelength smaller than $L$.
Conformal invariance protects the symmetries of the $M$ 
sector at short distances \cite{ls}, 
and any (super)symmetry breaking effect is only induced at 
distances larger than $L$. 
We can determine more quantitatively this decoupling
by using the AdS/CFT correspondence.
For example, we can calculate the amplitude
of the deep-inelastic scattering 
between a ``probe''particle, $e$, and a  CFT meson $M$,
mediated by $\Phi$ which we take to be a photon (see Fig.~4)
\begin{equation}
eM\rightarrow eX\, ,
\end{equation}
where by $X$ we refer to any combination of final states.
In the  5D AdS picture, the photon
is simultaneously coupled to the boundary at $y=\pi$  
(where $M$ lives), and to 
the boundary at $y=0$ (where non-CFT fields, like $e$, live).
Therefore, the amplitude is proportional to the 5D photon
propagator $G(p,y=0,y^\prime=\pi)$ calculated in Ref.~\cite{pomarol}
where $p$ is the 4D momentum.
This propagator drops exponentially 
at momentum scales larger than $ka(\pi)=1/L$,
\begin{equation}
G(p,y=0,y^\prime=\pi)\sim e^{-p L}\, ,
\end{equation}
and shows that the CFT meson $M$ quickly decouples from the photon 
when $pL>1$. Alternatively at high momenta, 
where the conformal symmetry is restored, the meson constituents
are no longer localized particle states, 
and so become transparent to the short wavelength photon probe.

\begin{figure}[t]
        \begin{center}
        \begin{picture}(270,130)(0,0)
     \Text(45,100)[r]{$e$}
            \Text(180,120)[l]{$e$}
           \Text(60,29)[r]{$M$}
            \Text(200,29)[l]{$X$} 
            \Text(130,70)[l]{$\gamma$} 
        \ArrowLine(55,100)(100,100)
        \ArrowLine(100,100)(170,120)

         \Photon(100,100)(135,30){3}{7} 

         \Line(65,31)(135,31)
         \Line(65,28)(135,28)
         \ArrowLine(140,29.5)(190,29.5)
         \ArrowLine(140,34)(190,44)
         \ArrowLine(140,25)(190,15)

         \Vertex(100,100){2.5}
         \Vertex(135,29.5){8}

     \end{picture}
\caption{\it Inelastic scattering of a fundamental field, $e$, with
a CFT bound-state, $M$.}
\end{center}
\end{figure}
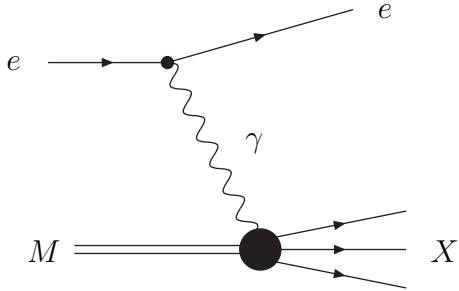

We must stress that this is not a general property of composite models.
For example, a theory like technicolor where the strong dynamics are 
assumed to be similar to QCD will not show this decoupling.
At high energies, the non-supersymmetric sector is not exponentially 
decoupled, since it can  couple to the constituents of the  bound-states. 
Therefore bound-states made of scalars (partners of the techniquarks
in a supersymmetric technicolor sector)
will receive large corrections to their masses.

\section{The SM partly supersymmetric}

Here we want to consider the possibility that only the Higgs sector 
of the SM is supersymmetric. Our motivation is to obtain a Higgs 
mass parameter that is insensitive to high-energy physics, and is induced
at low energies at the quantum level. This will guarantee a partial  
decoupling between the scale of new physics and the electroweak scale that
experiments, in particular LEP, are suggesting.
From the previous section, we know that this can be achieved 
by requiring the Higgs to be a CFT bound-state, while the SM fields 
are fundamental. This  scenario is most simply realized if we start 
in the AdS picture, where a slice of AdS$_5$ is bounded by a UV-brane 
with $\Lambda\sim M_P$ (or Planck-brane), and an IR-brane with 
$\Lambda_{IR}=a(\pi)\Lambda\sim$ TeV (or TeV-brane).
We will also assume that $k\sim M_P$, and then $L=1/(a(\pi)k)\sim 1/$TeV.  
All the SM fields are assumed to reside in the bulk, except for the Higgs 
sector that will be localized on the TeV-brane.
We will start with a supersymmetric bulk theory, and then we will break 
supersymmetry on the Planck-brane at the Planck scale. 
The supersymmetry 
breaking on the Planck-brane will be parametrized by the spurion 
superfield, $\eta=\theta^2 F$.
Only bulk fields with Neumann boundary conditions 
can couple to the spurion superfield. 
Similar scenarios but with the breaking of supersymmetry
on the TeV-brane have been previously considered in 
Refs.~\cite{gp1,gp2,mp,WarpedMSSM}. 
Unlike the model that we will present here, 
these scenarios resemble the MSSM at low energies.

Consider first the gauge sector \cite{gp1}, where the supersymmetric action in 
superfield notation can be found in Ref.~\cite{mp}. The 4D massless 
spectrum corresponds to an $N=1$ vector multiplet
\begin{equation}
V=\{A_\mu^a, \lambda^a,  D^a\}\, .
\end{equation}
It has a wave-function that is flat, and couples with equal strength 
to the two branes. 
Therefore, by adding the extra term
\begin{equation}
\int d^2\theta\,\frac{\eta}{\Lambda^2} \frac{1}{g^2_5}WW\,\delta(y)+h.c.\, ,
\label{sbg}
\end{equation}
the gaugino $\lambda^a$ will receive a huge mass $m_\lambda\sim 
F/\Lambda\sim M_P$ (see Appendix). 
Consequently, the gaugino effectively decouples, 
and the spectrum reduces to simply the gauge boson
$A_\mu^a$, and the auxiliary field $D^a$ with the Lagrangian
\begin{equation}
{\cal L}=-\frac{1}{4g^2}F^{a \mu\nu}F_{\mu\nu}^a+\frac{1}{2g^2} D^a D^a\, ,
\label{lagg}
\end{equation}
where $g$ is the 4D gauge coupling related to the 5D coupling by
\begin{equation}
  \frac{1}{g^2} = \frac{\pi R}{g^2_5}\, .
\label{gauge4D}
\end{equation}
Thus, at the massless level supersymmetry is completely broken, and
we just have the SM gauge bosons.

Similarly, matter fields (such as quarks and leptons) arise 
from 5D hypermultiplets. In the supersymmetric limit, the 4D massless spectrum 
is described by an $N=1$ chiral multiplet~\cite{mp}
\begin{equation}
    Q=\{\tilde q, q,  F_Q\}\, .
\end{equation}
Their wave-functions depend on the 5D hypermultiplet mass, and 
for simplicity we will assume the value $M_{5D}=k/2$ ($c=1/2$ in the 
notation of Ref.~\cite{gp1}) that corresponds to flat 
wave-functions~\cite{gp1,mp}. For this value of $M_{5D}$
the holographic interpretation is the same as that 
for the gauge sector (see Appendix).
Other values of $M_{5D}$ will be considered for the top quark
in Section~\ref{effpot}. Supersymmetry breaking is induced by adding 
on the Planck-brane the interaction
\begin{equation}
  - \int d^4\theta\,\frac{\eta^\dagger\eta}{\Lambda^4}\,Q^\dagger Q\,
  k \delta(y)\, ,
\label{sbm}
\end{equation}
which gives rise to squark and slepton masses of order 
$m_{\tilde q}\sim F/\Lambda\sim M_P$ (see Appendix). 
These scalar superpartners effectively decouple, and the 
massless spectrum solely consists of the fermions $q$ and auxiliary
fields $F_Q$.
 
In this way all the supersymmetric partners of the SM fields have
received Planck-scale masses, and the massless spectrum reduces to the usual
SM. At the massive level, the first KK-states appear at the scale 
$1/L\sim$ TeV. Unlike the massless states, the KK-spectrum is 
(approximately) supersymmetric. This is because the wave-functions of 
the KK-states are localized toward the TeV-brane and are less 
sensitive 
to the supersymmetry breaking on the Planck-brane.

The theory considered so far is the 5D dual theory of the 
SM coupled to a supersymmetric CFT theory.
Next we will consider the Higgs sector. Since we want the Higgs sector 
to be supersymmetric, it must be confined to the TeV-brane. 
In the 4D dual, this corresponds to having a CFT composite Higgs. 
The supersymmetric partner of the Higgs, the Higgsino, generates gauge
anomalies that must be cancelled. This leads us to consider the following
three possible Higgs scenarios:
\begin{list}{}{}
\item {\bf a) Two Higgs doublets:}  As in the MSSM, an extra Higgs 
(and Higgsino) is introduced to cancel the anomalies.

\item {\bf b) One Higgs doublet:} The Higgsino anomalies are cancelled
by an extra fermion that, as for the SM fermions, arises
from a 5D hypermultiplet.
Hence, the scalar partner of the 
extra fermion will obtain a Planck-sized mass.

\item {\bf c) Higgs as a slepton:} No Higgsino is introduced. Instead,
the tau (or other lepton) is assumed to be the supersymmetric
partner of the Higgs. For this purpose the tau must live on the 
TeV-brane (contrary to the rest of the matter fields).
\end{list}

\noindent
Let us now discuss in detail each of these possibilities.

\subsection*{a) Two Higgs doublet model}
Suppose that the Higgs sector consists of two $N=1$ 4D chiral 
multiplets, $H_1=\{h_1,\tilde h_1,F_{H_1}\}$, and 
$H_2=\{h_2,\tilde h_2,F_{H_2}\}$. 
The fact that the Higgs sector is doubled
guarantees two things. First, anomalous contributions arising from the 
Higgsinos $\tilde h_i$ are automatically cancelled, and secondly, 
that quark and lepton masses can be simply obtained from the 
following superpotential
on the TeV-brane
\begin{equation}
    \int d^2\theta\, \big[y_dH_1QD+y_uH_2QU+y_eH_1LE\big]\, .
\label{lagy}
\end{equation}
Although the Higgs spectrum is supersymmetric, it directly couples to 
the SM which does not respect supersymmetry (since it is broken at the 
Planck scale). This gives rise to the following effective theory 
for the Higgs
\begin{equation}
   {\cal L}=-|D_\mu h_i|^2-i{\bar{\tilde h}}_i \dsl \tilde h_i
    -D^a (h_i^\dagger T^a h_i)+
    y_dh_1qd+y_eh_1le+y_uh_2qu\, ,
\label{lagh}
\end{equation}
where $D_\mu$ is the covariant derivative, and $T^a$ are the generators
of the gauge group. 
The complete effective Lagrangian below the TeV scale is 
given by Eqs.~(\ref{lagg}) and (\ref{lagh}).
Eliminating the auxiliary field $D^a$, we obtain the 
tree-level Higgs potential
\begin{equation}
{ V}=\frac{1}{2g^2}D^aD^a=\frac{g^2}{2}(h_i^\dagger T^ah_i)^2\, .
\label{poth}
\end{equation}
For the neutral component of  the Higgs the  potential
becomes
\begin{equation}
    V=\frac{1}{8}(g^2+g^{\prime 2})(|h_1|^2-|h_2|^2)^2\, ,
\label{poth2}
\end{equation}
where $g$ and $g^\prime$ are the SU(2)$_L$ and U(1)$_Y$
couplings, respectively.
We have obtained an interesting result. Although the Higgs mass 
is zero by supersymmetry, the Higgs potential has a quartic coupling
that does not respect supersymmetry since it is 
generated from the $D$-term of the gauge sector.
At the quantum level, radiative corrections will induce 
a mass term for the Higgs which will be naturally smaller than $1/L$. 
In section~\ref{effpot} we will calculate the one-loop effective 
potential. For now we will present the result of the Higgs mass induced by 
gauge loops given by the two-point contribution to the effective potential
(first term of the sum of Eq.~(\ref{effpotg}))
\begin{equation}
     m_{h_i}^2 \simeq  \left(\frac{0.14}{L}\right)^2\, ,
\label{hmass}
\end{equation}
where $1/L=a(\pi)k\sim$ TeV.  
This is a finite contribution since it comes from a non-local effect 
(see Fig.~1). As expected, the result (\ref{hmass}) shows that
the supersymmetry breaking effects in the Higgs sector are an order of 
magnitude below the scale $L^{-1}$, and are much smaller than the Planck scale.
This gauge contribution is positive but, as we will show in 
Section~\ref{effpot}, can be overcome by a negative contribution from 
the top-quark loop, in order to trigger electroweak symmetry breaking.

\subsubsection*{Higgsino masses: The $\mu$-problem}

While radiative corrections  induce a scalar Higgs mass, 
the Higgsinos remain massless. Of course, phenomenologically
this is a problem, and we need to simultaneously consider possible 
mechanisms for generating Higgsino masses. In addition, the minimum of 
the Higgs potential depends critically on the value of
the $B\mu$-term (the bilinear term $B\mu h_1h_2$). 
This term is not generated by radiative corrections.
This means that while 
$\langle h_2\rangle \neq 0$, we have that $\langle h_1\rangle = 0$, 
and the fermions in the down-sector remain massless. So, we will 
also need to generate a VEV for $h_1$.

The simplest possibility is to consider the following superpotential 
on the TeV-brane 
\begin{equation}
    \int d^2\theta\,\, \left[\mu H_1 H_2 + \frac{\beta}{2\Lambda_{IR}} 
     (H_1 H_2)^2\right]\, .
\label{mupot}
\end{equation}
The first term directly gives a mass to the Higgsino. By supersymmetry
the Higgs scalar field will also receive a mass, since
Eq.~(\ref{mupot}) leads to the Higgs potential
\begin{equation}
   V=\Big[|\mu|^2+\frac{\beta\mu^\ast}{\Lambda_{IR}} 
(h_1h_2+h.c.)\Big] (|h_1|^2+|h_2|^2)+
{\cal O}\left(\frac{h^6}{\Lambda^2_{IR}}\right)\, .
\label{potthdm}
\end{equation}
In order to have electroweak symmetry breaking, the value of 
$|\mu|^2$ cannot be larger than the negative one-loop top-quark contribution.
This requires that $\mu\sim m_h\sim 0.1/L$ 
(if this value is smaller, then the Higgsino mass will be too small).
The second term of Eq.~(\ref{mupot}) has been introduced to generate a linear 
term in $h_1$ (it plays the role of the $B\mu$-term in the MSSM).
When  $h_2$ gets a VEV, this linear term generates a VEV for $h_1$ 
of order
$\langle h_1\rangle\sim -\beta\langle h_2\rangle^3/
(\mu\Lambda_{IR})\sim 0.1\langle h_2\rangle$ .

Instead of introducing a $\mu$-term from the beginning in our theory,
we can imagine generating it by the one-loop supersymmetry breaking
 effects. In this case the $\mu$-term 
will naturally be of order the electroweak scale.
For example, we can consider a sector that has a field $X$ whose
$F$-term  squared is induced at the one-loop level.
If this field couples to the Higgs in the following way
\begin{equation}
        \int d^4\theta\, \frac{X^\dagger}{\Lambda_{IR}} H_1 H_2~,
\end{equation}
then it will generate a $\mu$-term with $\mu=\langle F_X\rangle/\Lambda_{IR}$. 
Furthermore, $h_1$ does not need to get a VEV since 
fermion masses can be generated from the Kahler potential terms
\begin{equation}
    \int d^4\theta\, \frac{X^\dagger}{\Lambda_{IR}^2} (H_2^\dagger QD 
     + H_2^\dagger LE)~.
\end{equation}
The bottom quark Yukawa coupling will be given by 
$y_b\sim \langle F_X\rangle/\Lambda_{IR}^2\sim 0.1$, 
where we have assumed that the Kahler-term coupling is of order one. 
For the other quarks and leptons in the down sector,
these couplings will need to be hierarchically smaller.

It is also possible to introduce a singlet chiral supermultiplet, $S$, on the
TeV-brane with a superpotential $S H_1 H_2+S^3$. 
A  $\mu$-term can then be generated if $S$ gets a VEV
that, parametrically, will be of the same order as the electroweak scale.
This possibility deserves further analysis which will not be carried
out here.

\subsection*{b) One Higgs doublet model}

A natural solution to the $\mu$-problem exists if the Higgs arises from 
a 5D bulk hypermultiplet that consists of two $N=1$ chiral multiplets,
$H_{1,2}$, of opposite charges.
If the boundary conditions for $H_2$ are taken to be Neumann, while 
those for $H_1$ are Dirichlet (as in the matter sector), then 
the 4D massless sector will correspond to a single 4D chiral multiplet. 
This theory will then be anomalous. 

However, if $H_1$ is assumed to have Neumann boundary conditions on the 
Planck-brane but Dirichlet on the TeV-brane (and vice-versa for $H_2$),
then the lowest lying state is a massive pair of 4D chiral multiplets.
These twisted boundary conditions then lead to a theory that is not anomalous.
The mass of these chiral multiplets can be written as a superpotential term,
$\mu H_1H_2$, where the value of $\mu$ depends on the 5D hypermultiplet mass.
In particular, assuming a 5D mass term $M_{5D}=k/2$, we find that
\footnote{For $M_{5D} < k/2$ the two Higgsinos are localized
towards the TeV-brane and the $\mu$-term becomes ${\cal O}$(TeV).
For  $M_{5D}>k/2$ one Higgsino  becomes
strongly localized towards the Planck-brane, while the other towards 
the TeV-brane. In this case
the $\mu$-term is driven to exponentially small values,
$\mu\sim e^{-(M_{5D}/k-\frac{1}{2})\pi k R}$.}
\begin{equation}
\label{muterm}
    \mu \simeq \sqrt{\frac{2}{\pi k R}} k e^{-\pi kR}~.
\end{equation}
This mass term is analogous to the gaugino mass term obtained 
in Ref.~\cite{gp2}. Note that the $\mu$-term is naturally suppressed 
below the TeV-scale by the factor, $1/\sqrt{\pi kR}$. 
The Higgs wave-functions of the lowest lying 
modes become
\begin{eqnarray}
    {H}_2(y) &\simeq& \sqrt{2\pi k R} e^{-\pi k R} \frac{|y|}{\pi R}
     e^{\frac{5}{2}k |y|}~,\\
   {H}_1(y) &\simeq& 2 e^{-\pi k R} \sinh[k(|y|-\pi R)]
   e^{\frac{5}{2}k |y|}~,
\end{eqnarray}
showing that ${H}_2$ is localized towards the TeV-brane,
while ${H}_1$ is localized towards the Planck-brane.
Hence, $H_1$ will be sensitive to the breaking of supersymmetry, 
and as in the matter sector, the scalar $h_1$ will receive a Planck mass.

Thus, the effective theory of the bulk Higgs with twisted boundary conditions
below the TeV-scale consists of one chiral supermultiplet $H_2$, 
  an extra fermion $\widetilde h_1$, and the $F_{H_1}$ auxiliary field.
The effective Lagrangian is given by
\begin{equation}
\label{effth}
     -{\cal L}_{eff} = \mu {\widetilde h_1}{\widetilde h_2} + 
     \mu^2 |h_2|^2 + \frac{1}{8}(g^2+g'^2) |h_2|^4\, ,
\end{equation}
where the second (last)  term in (\ref{effth}) is the 
$F_{H_1}$-term ($D$-term)  contribution. 
In the 4D dual description, $h_2$ and $\widetilde h_2$ are composite 
states of the CFT, while $\widetilde h_1$ is a fundamental field that
has  been added to the CFT. The $\mu$-term can then be understood as arising
from the marriage of the Higgsino ${\widetilde h}_1$ with the fermion 
bound-state corresponding to ${\widetilde h}_2$. This is exactly
analogous to the dual interpretation of the gaugino mass 
in the warped MSSM~\cite{gp2}.

\subsection*{c) Higgs as a slepton}

Neither anomalies nor the $\mu$-problem will arise if the 
Higgs is considered to be the superpartner of the tau (or other
lepton), and forms a 4D chiral multiplet, $L_3=(h,\tau,F_\tau)$, on 
the TeV-brane. This idea is not new, and dates back to the early days of 
supersymmetry~\cite{fayet}. The major obstacle in implementing this 
identification is that neutrino masses are large, and consequently, 
experimentally ruled out. This is because the gaugino induces the 
effective operator
\begin{equation}
\label{neutop}
         \frac{g^2}{M_\lambda}\nu \nu {h} {h}~,
\end{equation}
where $M_\lambda$ is the gaugino mass. Thus, 
for $M_\lambda \sim \rm TeV$, this operator generates a 
neutrino mass of order $\langle h\rangle^2/M_\lambda\sim 10$ GeV.

However, in our warped model with only a partly supersymmetric spectrum,
there is an approximate $U(1)_R$ symmetry which acts as a continuous 
lepton symmetry, and suppresses the neutrino masses.
This symmetry is exact in the low-energy effective 
Lagrangian because there is no gaugino partner 
to the SM gauge boson, and the Kaluza-Klein gauginos have Dirac masses 
which are invariant under the $U(1)_R$. 
However, the $U(1)_R$ symmetry is  broken at a high scale by
the large gaugino mass $M_\lambda\sim M_P$,
naturally giving small neutrino masses  $m_\nu\sim 10^{-5}$ eV.

Having obtained acceptable neutrino masses, we can also generate the
required fermion mass spectrum from operators only involving $L_3$. The 
couplings that generate masses for the down quarks and charged leptons 
can come from the superpotential (on the TeV-brane)
\begin{equation}
      \int d^2\theta\, \left(y_d^{(i)}  L_3 Q_i D_i+ y_e^{(i)} L_3 L_i E_i
      \right)~,
\end{equation}
except for the third generation charged fermion in $L_3$, since
due to the antisymmetry of the $SU(2)$ indices, $L_3 L_3=0$.
In this case one must rely on higher-dimensional operators in the Kahler 
potential~\cite{gk}.
By holomorphy, there is no superpotential term which generates masses for 
the up quarks. Instead one must consider Kahler potential couplings like
\begin{equation}
     \int d^4\theta\, y_u^{(i)} 
\frac{X^\dagger}{\Lambda_{IR}^2} L_3^\dagger 
      Q_i U_i~,
\label{ktop}
\end{equation}
where $X$ is a spurion field that, as in the previous examples, 
has an $F$-term, $F_X$, on the TeV-brane. Notice, however,
that one needs $F_X\lappeq 0.1\Lambda_{IR}^2$, otherwise 
Kahler terms such as $X^\dagger X L_3^\dagger L_3$ 
give a large contribution to the Higgs mass-parameter, and
lead to an electroweak scale which is too close to $\Lambda_{IR}$.
Such a  small value of $F_X$ can be 
problematic to generate the  top quark mass from
Eq.~(\ref{ktop}) unless the coefficient $y_u^{(3)}$ is large.
In spite of this problem, 
we find this scenario very interesting because no extra matter is needed 
in order to have supersymmetry protect the Higgs mass. 

\subsection{One-loop effective potential and electroweak symmetry breaking}
\label{effpot}

The  Higgs fields will receive supersymmetry-breaking contributions
from the bulk vector multiplets and bulk hypermultiplets containing 
the quarks and leptons  as shown in Fig.~1.
The simplest way to compute these contributions 
is to use the 5D AdS propagator in loop calculations. 
The general expressions for these propagators in a slice of AdS$_5$ 
were presented in Ref.~\cite{gp2}. Since the Higgs multiplet is
located on the TeV-brane, we will be interested in the propagator expressions
evaluated at the TeV boundary ($y=y^\prime=\pi$).
Following the notation of Ref.~\cite{gp2}, we have that
for bulk fields $\{V_\mu, \phi, e^{-2\sigma}\psi_{L,R}\}$
with  masses ${\widehat M}^2=\{0, ak^2,c(c\pm1)k^2\}$, the general
expression for the propagator is given by
\begin{equation}
\label{gengf}
      G(p) =-\frac{e^{s\pi kR}}{k} \left[
       \frac{ {J}^P_\alpha(ip/k) Y_\alpha(ip e^{\pi kR}/k)
       - Y^{P}_\alpha (ip/k) J_\alpha(ipe^{\pi kR}/k) }
       {J^{T}_\alpha(ipe^{\pi kR}/k) Y^{P}_\alpha(ip/k)
     - Y^{T}_\alpha(ip e^{\pi kR}/k) J^{P}_\alpha(ip/k)}
        \right]~,
\end{equation}
where $\alpha=\sqrt{(s/2)^2+{\widehat M}^2/k^2}$, $s=\{2,4,1\}$  and
\begin{equation}
     J^{i}_\alpha(x) = 
(\frac{s}{2}-\alpha- {r_i}) J_\alpha(x)+x J_{\alpha-1}(x)\, ,\ \ 
i=P,T\, .
\end{equation}
The functions $J_\alpha$ and $Y_\alpha$ are Bessel functions,
and the values of $r_P (r_T)$ are determined by
the boundary conditions on the Planck (TeV) brane.
Only for fields with Neumann boundary conditions on the TeV-brane will
the propagator Eq.~(\ref{gengf}) be nonzero.

\subsubsection{Bulk gauge contributions}

The gauge contributions to the Higgs potential arise from loops of 
gauge bosons, gauginos and $D$-terms.
For the SU(2)$_L$ gauge sector the contribution
to the effective potential of the neutral component of the Higgs $h$
is given by
\begin{eqnarray}
     V_{gauge}(h) &=& 6 \sum_{n=1}^{\infty} \int_0^\infty 
      \frac{dp}{8\pi^2}\, p^3 \frac{(-1)^{n+1}}{n}
      \left[ G_B^n(p) - G_F^n(p) \right] m_W^{2n}(h)\nonumber\\
     &=&6 \int_0^\infty \frac{dp}{8\pi^2}\, p^3 
      \log\left[\frac{1+m_W^2(h) G_B(p)}{1+m_W^2(h) G_F(p)}\right]~,
\label{effpotg}
\end{eqnarray}
where $m_W^2(h)=g^2|h|^2/2$. The boson propagator is defined as
$G_B(p)=\pi R G(p)$, where $G(p)$ is given by Eq.~(\ref{gengf})
with $\alpha=1$, and $r_T=r_P=0$, while for the fermion propagator
we define $G_F(p)=e^{\pi k R} \pi R G(p)$, where $G(p)$
is obtained with $\alpha=1$, 
$r_P=-\frac{1}{2}+\frac{ip}{4k}\frac{F}{\Lambda^2}$, and 
$r_T=-\frac{1}{2}$. The breaking of supersymmetry is  parametrized by $F$.
We must stress that the effective potential 
is very insensitive to the actual value 
of $F$ since the contribution to the integral in Eq.~(\ref{effpotg})  comes
from the region $p^2\approx  1/L^2\ll F$.
The gauge contribution generates a potential that monotonically 
increases with $h$.

\subsubsection{Bulk matter contributions}

Similarly, we can calculate the contribution from the bulk  
hypermultiplets to the Higgs effective potential. Unlike the gauge 
contributions, the loop diagrams now involve two bulk fermions. 
For the top quark the contribution to the effective potential is 
given by
\begin{eqnarray}
     V_{top}(h) &=& 6
      \sum_{n=1}^{\infty} \int_0^\infty 
      \frac{dp}{8\pi^2}\, p^{2n+3} \frac{(-1)^{n+1}}{n}
      \left[G_B^{2n}(p) - G_F^{2n}(p) \right] m_t^{2n}(h)\nonumber\\
     &=&6\int_0^\infty \frac{dp}{8\pi^2}\, p^3 
      \log\left[
\frac{1+p^2 m_t^{2}(h) G_B^2(p)}{1+p^2 m_t^{2}(h) G_F^2(p)}\right]~.
\label{effpott}
\end{eqnarray}
where $m_t^2(h)=y_t^2 |h|^2$, with $y_t$ defined as the 4D top-quark
Yukawa coupling. The scalar boson propagator is given by $G_B(p)=
 e^{-2\pi kR} G(p)/f^2_t$ where $f_t$ is the fermion zero-mode wave-function
evaluated at $y=\pi $, and
$G(p)$ is given by Eq.~(\ref{gengf}) with $\alpha=|c_t+\frac{1}{2}|$, 
$r_P=\frac{3}{2}-c_t+\frac{F^2}{2\Lambda^4}$ and $r_T=\frac{3}{2}-c_t$.
Note that we are assuming an arbitrary 5D mass, $M_{5D}=c_t k$.
For the fermion we instead have $G_F(p)= e^{\pi kR} G(p)/f^2_t$ 
where $\alpha=|c_t+\frac{1}{2}|$, and $r_P=r_T=-c_t$.
The top quark contribution generates a potential that monotonically decreases
with $h$, thus making it possible to break the electroweak symmetry.

\subsubsection{Electroweak symmetry breaking and the physical Higgs mass}
\label{313}

To study electroweak symmetry breaking (EWSB),  
we will restrict to a single Higgs doublet. 
The effective potential of the neutral component, $h$ is given by
\begin{equation}
  V(h)=\mu^2|h|^2+\frac{1}{8}(g^2+g'^2)|h|^4 + V_{gauge}(h)+ V_{top}(h)\, .
\label{higgspot}
\end{equation}
This potential corresponds to the Higgs of model (b), and, 
for $\mu=0$, to that of model (c).
Although model (a) has another Higgs $h_1$,  its effect
on the breaking of electroweak symmetry is small since, as was 
argued earlier, $h_1$ obtains a VEV that is smaller than that of $h_2$.
Note that we are only considering the one-loop contributions arising 
from  the SU(2)$_L$ gauge sector Eq.~(\ref{effpotg}) and the 
top-quark sector Eq.~(\ref{effpott}).

The potential Eq.~(\ref{higgspot}) depends on two parameters,
$\mu$ and  $c_t$. 
For $\mu$ close to zero, we find that $c_t\gappeq -0.5$ in order
to have EWSB.  
In particular, for  $c_t\simeq -0.5$ 
we obtain the prediction $L^{-1}\simeq 2$ TeV.
In this case, however, we find a very light Higgs
$m_{\rm Higgs}\simeq 95$ GeV.
By increasing $c_t$, we increase the top-quark contribution to the 
effective potential. The Higgs mass can then be larger, but the scale $L^{-1}$
becomes closer to the electroweak scale.
For example, when $c_t\simeq -0.4$, we have $L^{-1}\simeq 350$ GeV and 
$m_{\rm Higgs}\simeq 100$ GeV.
If $\mu$ is nonzero, we have more freedom  and a  heavier Higgs
can be obtained.
For $\mu\simeq 200$ GeV and $c_t\simeq -0.4$ we find $L^{-1}\simeq 1$ TeV and 
$m_{\rm Higgs}\simeq 105$ GeV.
Larger values of $L^{-1}$  and the Higgs mass can be obtained,
but this requires  a precise tuning between $c_t$ and $\mu$.
The fact that $c_t$ is preferred to be approximately $-0.5$ (instead of 
near $0.5$ as assumed for the other quarks) implies that, 
in the 4D dual theory, the top quark is mostly a CFT bound-state instead 
of a fundamental state (see Appendix). The stop is then present in the 
low-energy spectrum with a mass of order TeV. All these results, however,  
are subject to  some uncertainties which we will discuss next.

There are other one-loop corrections
to the Higgs potential which we did not consider in Eq.~(\ref{higgspot}).
An important one-loop effect is the  renormalization of the $D$-term in 
Eq.~(\ref{poth}). The $D$-term is proportional to the gauge coupling, 
and therefore the renormalization of the $D$-term depends on the 
corrections of the gauge coupling.
In a slice of AdS$_5$ the gauge coupling receives logarithmic
corrections at the one-loop level due to the fundamental fields
(zero modes)~\cite{pomarol}.
This makes the gauge couplings in Eq.~(\ref{poth2})
differ from the gauge couplings measured at low-energies
by sizable logarithmic corrections. 
In the case where  the fundamental sector consists of only the SM,
these corrections reduce the Higgs quartic coupling by $\sim 10\%$.
The origin of these corrections is easily understood in the 4D dual theory.
By supersymmetry, the Higgs quartic coupling is proportional to the gauge
coupling. However, in our scenario supersymmetry is broken in the 
gauge sector at high energies $F/\Lambda\sim M_P$. Hence
the evolution from  $M_P$ to TeV of the gauge couplings is different 
compared to that of the Higgs quartic coupling, because
the two kinetic terms in (\ref{lagg}) do not have the same
renormalization.
Since supersymmetry is only broken in the fundamental sector
only these fields can contribute to this  difference.

Other type of effects that can affect the electroweak breaking and 
the Higgs mass are due to boundary terms that can be present in the theory.
For example, we can have a term like
\begin{equation}
       - \int d^4x \int dy\,\,  
     \frac{1}{4g_b^2} F^{a\mu\nu} F_{\mu\nu}^a \delta(y-\pi)\, .
\label{bkt}
\end{equation}
Eq.~(\ref{bkt}) then modifies Eq.~(\ref{gauge4D}) to
\begin{equation}
    \frac{1}{g^2}=\frac{\pi R}{g_5^2} + \frac{1}{g_b^2}\, .
\label{gaugec}
\end{equation}
Also the boundary conditions for the propagators are now modified due to
the presence of the boundary kinetic term.
Their effect can easily be incorporated into the propagators
of Eq.~(\ref{gengf}) by taking
\begin{equation}
        r_T= \frac{g_5^2 p^2}{2g^2_bk} 
e^{2\pi kR}\equiv\epsilon\, \frac{p^2}{k^2} e^{2\pi kR} \, .
\label{eps}
\end{equation}
For positive values of $\epsilon$ the gauge contributions to the effective
potential become smaller. For example, when 
$\epsilon\simeq 1$, Eq.~(\ref{hmass}) changes to 
$m_{h_i}^2 \simeq  \left(0.08/L\right)^2$ and the  physical Higgs boson 
mass can increase by approximately 10 GeV. Thus we see that this type of 
boundary effect can be a substantial correction to the Higgs mass.

Also higher-dimensional operators can contribute to the Higgs mass.
For example, if the  $\mu$-term is nonzero, 
the operator of Eq.~(\ref{eff}) gives a contribution to 
the Higgs quartic coupling.
However, the coefficient of the  operator of Eq.~(\ref{eff})  
cannot be very large, otherwise it will also give a very 
large correction to the electroweak observables, which we will
comment on in the next section.

In summary, due to the above uncertainties, we cannot obtain
a precise prediction for the scale $L$ as a function of the electroweak 
scale. Nevertheless, we can conclude that without any large tuning of the
parameters, the scale of new physics, $L^{-1}$ lies an order of magnitude 
above the electroweak scale, and the physical Higgs mass is smaller 
than approximately 120 GeV.

\subsection{Electroweak bounds}
\label{ewb}
The success of the SM predictions places strong constraints on 
the scale of new physics. In our model the KK-states affect the 
SM relations between the electroweak observables. There are two 
kinds of effects \cite{mapo}. The first effect arises from the exchange 
of the KK excitations of the $W$, $Z$ and $\gamma$, which 
induces extra contributions to four-fermion interactions.
However, these contributions are very model dependent, and can be zero
for certain cases where the SM fermions do not couple to the 
KK-states~\cite{gp1}. Therefore, we will not consider them here.
A second type of effect arises if the Higgs is on the boundary.
In this case the SM gauge bosons will mix with the KK-states,  and 
modify their masses \cite{mapo}. 
These are the most important effects, and we will study them below.
A similar analysis can also be found in Ref.~\cite{rsew}.

In superfield notation the 4D Lagrangian of the SM gauge-boson KK-states 
is given by
\begin{equation} 
   {\cal L}=\int d^4\theta\sum_n\Big[\left. H_i^\dagger e^{-2g_5[f_0V
     +f_{n}V^{(n)}]}H_i\right|_{y=\pi}+M_{n}^2 V^{(n)\, 2}\Big]\, ,
\end{equation}
where $f_0$ ($f_n$) is the wave-function of the massless mode $V$
(KK-state $V^{(n)}$), and $i$ labels the number of SM Higgs doublets.
Integrating out $V^{(n)}$, by using their equation of motion,
we obtain the effective 4D Lagrangian term
\begin{equation} 
{\cal L}_{eff}=-\int d^4\theta\sum_n\frac{g^2}{M^2_{n}}
\frac{f^2_n(\pi)}{f^2_0(\pi)}(H_i^\dagger e^{-2gV}T^aH_i)^2\, ,
\label{eff}
\end{equation}
where $g$ is given in Eq.~(\ref{gauge4D}).
When the Higgs boson obtains a VEV, this operator will induce extra
mass terms for the SM gauge bosons, $W_\mu$ and $Z_\mu$, namely
\begin{equation} 
  -\frac{1}{2} Xm^2_Z Z_\mu Z^\mu-\frac{1}{2} X\frac{m^4_W}{m_Z^2} 
   W_\mu W^\mu\, ,
\end{equation}
where in analogy with the flat extra dimension case \cite{dpq} 
we have defined
\begin{equation} 
X=\sum_n\frac{m^2_Z}{M^2_{n}}\frac{f_n^2(\pi)}{f^2_0(\pi)}\, .
\end{equation}
The value of $X$ can easily be calculated from the 5D gauge boson propagator
Eq.~(\ref{gengf})
by subtracting out the the zero-mode contribution. Thus, we obtain
\begin{equation}
X=\frac{m^2_Z}{f^2_0(\pi)}\left[G(p=0)
-\frac{f^2_0(\pi)}{p^2}\right]
\simeq\left(4.1\, m_Z L \right)^2\, .
\label{xdef}
\end{equation}
By comparing with the flat extra dimension case \cite{dpq}, where 
$X\simeq (1.8\, m_Z R_{\rm flat})^2$, we obtain the following
bound for the warped case 
\begin{equation} 
R^{-1}_{\rm flat}\gappeq 4\ {\rm TeV}\ \ \ \
 \Rightarrow \ \ \ \
L^{-1} \gappeq 9\ {\rm TeV}\, .
\label{constrain}
\end{equation}
Although the bound Eq.~(\ref{constrain}) 
is quite strong, we must keep in mind that there 
are  inherent uncertainties in the above calculation. 
First, there can be brane kinetic terms (\ref{bkt}). 
These can be taken into account by using Eq.~(\ref{eps})
in the propagator part of Eq.~(\ref{xdef}).
However, we find that for $\epsilon\sim 1$ the bound 
Eq.~(\ref{constrain}) is not affected very much.
Second and more importantly,  we have used Eq.~(\ref{gauge4D})
to relate the  coupling of the KK-states to the Higgs 
(the $g^2$  in front of Eq.~(\ref{eff})) with 
the gauge coupling measured at  low-energy experiments.
However, Eq.~(\ref{gauge4D}) receives one-loop corrections 
$\propto\ln(k/{\rm TeV})=\pi kR$~\cite{pomarol} that are model dependent.
This gives a sizable uncertainty to the bound in Eq.~(\ref{constrain}).

\subsection{The TeV/$M_P$ hierarchy and radius stabilization}

In our model the 
large  hierarchy between the Planck scale and $1/L$ is 
explained by the warp factor $a(\pi)=e^{-\pi k R}\sim$ TeV$/M_P$
as in the Randall-Sundrum model \cite{rs}.
However, this requires a stabilization mechanism for the radion.
Several mechanisms~\cite{gw,rsm}  have been  proposed in the past. 
The mechanism of Ref.~\cite{gw} is not supersymmetric, 
while we find that those of Ref.~\cite{rsm} are not operative
in  our scenario.

Nevertheless,
it has been recently pointed out \cite{gapo} that the radius
can be stabilized by quantum effects (Casimir energy)
if certain fields (e.g. gauge bosons) are in the bulk.
This is a very simple solution to the 
stabilization of the hierarchy 
that can also be operative in our scenario. 
Furthermore, it does not affect the results presented above.

\section{Conclusion}

We have presented a novel class of particle models
 in which different sectors of the theory
have different  scales of supersymmetry breaking.
The model is based on a 5D theory compactified in a slice of AdS
as shown in Fig.~1.
It has a very interesting  holographic interpretation:
one sector is composed of CFT bound-states,
while the other sector consists 
of fundamental fields.
Even if the scale of supersymmetry breaking is large ($M_P$),
the CFT sector  is only sensitive to
supersymmetry breaking effects
of order $L^{-1}$ (TeV), where $L$ is the size of the bound-states.

We have applied our scenario to the SM in which only the
 Higgs sector  is supersymmetric, and depending on the value
of $c_t$, also the top quark. While the supersymmetric Higgs 
sector can either consist of one or two Higgs doublets, 
an interesting alternative involves identifying the Higgs as  
the partner of the tau lepton.
The  Higgs mass-parameter is protected by supersymmetry, which can be 
an order of magnitude smaller than $1/L$, and much smaller than  $M_P$.
Therefore this scenario not only solves the hierarchy problem but 
naturally explains the ``little'' 
hierarchy between the electroweak scale and the composite 
Higgs scale $1/L\sim$ TeV
in a novel way. This is one of the main points of the paper.

The model also has several interesting predictions.
First, the quartic Higgs coupling is fixed by supersymmetry, 
and as in the MSSM, leads to a light Higgs.
The precise value of the Higgs mass is very model dependent
(like in the MSSM!) but if no large tuning of parameters is imposed, 
we obtain $m_{\rm Higgs}\lappeq 120$ GeV.
Second, the only supersymmetric SM partner is the Higgsino (and
possibly the stop), with a mass around the electroweak scale 
(unless we choose the tau to be the partner of the Higgs).
Experimental searches for this Higgsino are very 
different from ordinary chargino searches due to the degeneracy 
of the charged and neutral Higgsino component~\cite{higgsino}.
Finally, at energies $L^{-1}\sim $ TeV, there are
plenty of new CFT resonances (KK-states) 
that approximately respect supersymmetry.
This leads to new and interesting possibilities for physics at 
the TeV scale.

\section*{Acknowledgements}
The work of TG was supported in part by 
a DOE grant DE-FG02-94ER40823 at the University of Minnesota.
The work of AP was  supported  in part by 
the MCyT and FEDER Research Project
FPA2002-00748 and DURSI Research Project 2001-SGR-00188.

\newpage

\section*{Appendix: Mass spectrum  of bulk fields and 
its holographic interpretation}

In this Appendix we will derive the 4D mass spectrum 
of the supersymmetric bulk fields (vector and hypermultiplet)
when   supersymmetry is broken on the Planck-brane by Eqs.~(\ref{sbg})
and (\ref{sbm}).
We will show that the 4D particle states split into two types:
those which are sensitive to the Planck-brane are to be associated with 
fundamental fields in the 4D dual theory, while those which are sensitive 
to the TeV-brane are to be associated with CFT bound-states.
Only the first type of states directly feel the breaking of supersymmetry.

If we consider the supersymmetry-breaking terms Eqs.~(\ref{sbg})
and (\ref{sbm}) then the boundary mass terms for the gauginos and 
squarks are
\begin{equation}
\label{boundterm}
    - \int d^4 x\int dy\,\, \Big[\frac{F}{g^2_5\Lambda^2}\lambda^a\lambda^a
       +\frac{F^2}{\Lambda^4} k |{\widetilde q}|^2\Big] \delta(y)\, .
\end{equation}
The 4D mass spectrum is obtained from the poles of the
propagators of Eq.~(\ref{gengf})
\begin{equation}
\label{poles}
   {J^{T}_\alpha(me^{\pi kR}/k) Y^{P}_\alpha(m/k)
     = Y^{T}_\alpha(m e^{\pi kR}/k) J^{P}_\alpha(m/k)}\, .
\end{equation}
For the gauginos we have $\alpha=1$,
$r_P=-\frac{1}{2}+\frac{m}{4k}\frac{F}{\Lambda^2}$ and  $r_T=-\frac{1}{2}$.
For these values, the r.h.s. of Eq.~(\ref{poles})  can be neglected for
any value of the supersymmetry-breaking parameter 
$F$, and the masses can be determined by the zeros of 
$J^{T}_\alpha(me^{\pi kR}/k) Y^{P}_\alpha(m/k)$. 
This allows one to separate the solutions into two types.
Those that depend on the boundary conditions at the TeV-brane,
and those that depend on the boundary conditions at the Planck-brane
\begin{eqnarray}
J^{T}_1(me^{\pi kR}/k)=0\ \ &\rightarrow&\ \     J_0(  me^{\pi kR}/k)=0\, ,
\label{kks}
\\
Y^{P}_1(m/k)=0\ \ &\rightarrow&\ \        
m\simeq -\frac{F}{4\Lambda^2}k\left(\log
      \frac{m}{2k}+\gamma_E\right)^{-1}\, ,
\label{zeromode}
\end{eqnarray}
where $\gamma_E$ is the Euler-Mascheroni constant.
Eq.~(\ref{kks})  determines the KK spectrum and 
it is  associated with the 4D CFT spectrum. Note that the KK spectrum
does not depend on $F$, and so the gauge boson and gaugino KK masses are the
same. Eq.~(\ref{zeromode}) corresponds to 
 the ``zero mode'' \cite{pomarol} which in the 4D dual picture
is associated with a fundamental field 
\footnote{More precisely, the fundamental states are identified with the poles
of the 5D propagator evaluated at the Planck-brane ($y=y^\prime=0$)
in the limit of $R\rightarrow\infty$.}
(this solution is valid if $m\lappeq k$).
This zero mode is the  partner of the SM gauge boson
which, as advertised, has a mass of order $F/\Lambda$.
When $F\rightarrow 0$, this gaugino becomes massless.
If we actually take into account the r.h.s. of Eq.~(\ref{poles}), then this
corresponds to considering the mixing between the CFT bound-states and 
fundamental fields (Fig.~2).
This mixing introduces a breaking of supersymmetry 
in the  CFT sector.

For the  squarks, the mass spectrum  is determined by Eq.~(\ref{poles}) 
with $\alpha=|c+\frac{1}{2}|$, 
 $r_P=\frac{3}{2}-c+\frac{F^2}{2\Lambda^4}$ and $r_T=\frac{3}{2}-c$,
where $c$ parametrizes the 5D hypermultiplet mass $M_{5D}=ck$.
The holographic  interpretation of the mass spectrum
depends on the values of $c$.
For $c\geq \frac{1}{2}$, we have a situation similar to the gauge sector
where  the r.h.s. of Eq.~(\ref{poles})  can be neglected 
and the masses are determined by
     \begin{eqnarray}
J^{T}_{c+\frac{1}{2}}(me^{\pi kR}/k)=0\ \ &\rightarrow&\ \     
J_{c-\frac{1}{2}}(  me^{\pi kR}/k)=0\, ,
\label{kksh}
\\
Y^{P}_{c+\frac{1}{2}}(m/k)=0\ \ &\rightarrow&\ \        
m^2\simeq (c-1/2)\frac{F^2}{\Lambda^4}k^2
\ \ \ \ \ \ \ \ \ \ \ \ \ \ \ \ \ \ (c>\frac{1}{2})\, ,\nonumber\\
&\rightarrow&\ \        
m^2\simeq -\frac{F^2}{2\Lambda^4}k^2\left(\log
      \frac{m}{2k}+\gamma_E\right)^{-1}\ \ \ \ (c=\frac{1}{2})\, .
\label{zeromodeh}
     \end{eqnarray}
The state whose mass is given by
Eq.~(\ref{zeromodeh}) is the one to be associated, in the 4D dual picture,
to the fundamental state (the partner of the quark).  
Its mass 
 is  of order $F/\Lambda\sim M_P$ 
(this is valid if $m\lappeq k$), showing that
supersymmetry in the fundamental sector is broken at the scale 
$M_P$.
For $-\frac{1}{2}< c < \frac{1}{2}$ the r.h.s. of Eq.~(\ref{poles}) 
cannot be neglected independently of the value of $F$, 
and the solutions to Eq.~(\ref{poles}) 
cannot be separated into those that depend on the TeV-brane and those
that depend on the Planck-brane.
This makes it difficult to identify the fundamental 
and CFT states.
This non-decoupling effect is due to the large mixing that exists
between the sources $\Phi$ and the CFT states. 
For $c\leq -\frac{1}{2}$  the r.h.s. of Eq.~(\ref{poles}) 
can be neglected again, and we find
\begin{eqnarray}
J^{T}_{|c+\frac{1}{2}|}(me^{\pi kR}/k)=0\ \ &\rightarrow&\ \     
J_{\frac{1}{2}-c}(  me^{\pi kR}/k)=0\, ,
\label{kksh2}
\\
Y^{P}_{|c+\frac{1}{2}|}(m/k)=0\ \ &\rightarrow&\ \        
{\rm No\ solution\ for\ }  m\lappeq k\, .
\label{zeromodeh2}
     \end{eqnarray}
Thus, in this case there is no fundamental state  whose mass
is proportional to $F$, which would become massless in the limit
$F\rightarrow 0$. However, there is an extra massless mode that can 
be found by looking at the pole of the full propagator Eq.~(\ref{gengf})
in the limit that the Planck-brane decouples ($k\rightarrow\infty$ with
$e^{\pi kR}/k$ fixed). This massless mode corresponds to a 
CFT bound-state. It picks a tree-level mass from its mixing with the 
fundamental field~\cite{gapo} (this can be seen by not neglecting the 
r.h.s. of Eq.~(\ref{poles})).
In the formal limit $c\rightarrow -\infty$, 
the spectrum consists of a single
supersymmetric field localized on the TeV-brane.

A similar analysis can be done by
using the standard holographic correspondence \cite{adscft} to 
calculate the two-point functions $\langle {\cal OO}\rangle$
where ${\cal O}$ is the CFT operator of Eq.~(\ref{sources}). 
In a theory with a TeV-brane $\langle {\cal OO}\rangle$ has poles 
corresponding to the CFT bound-states. One will find that the CFT spectrum
is determined by Eqs.~(\ref{kks}), (\ref{kksh}) and (\ref{kksh2}).
If the sources are dynamical, then their propagators are given by
$1/(p^2-\langle {\cal OO}\rangle)$, and the mass spectrum
is obtained from $p^2-\langle {\cal OO}\rangle=0$.


\newpage

\begin{thebibliography}{99}


\bibitem{luty}
For an alternative possibility see, M.~A.~Luty,
Phys.\ Rev.\ Lett.\  {\bf 89}, 141801 (2002)
[{\tt arXiv:hep-th/0205077}].


\bibitem{little}
N.~Arkani-Hamed, A.~G.~Cohen and H.~Georgi,
Phys.\ Lett.\ B {\bf 513}, 232 (2001)
[{\tt arXiv:hep-ph/0105239}].


\bibitem{rs} 
L.~Randall and R.~Sundrum,
Phys.\ Rev.\ Lett.\  {\bf 83}, 3370 (1999)
[{\tt arXiv:hep-ph/9905221}].


\bibitem{adscft}
J.~M.~Maldacena,
Adv.\ Theor.\ Math.\ Phys.\  {\bf 2}, 231 (1998)
[{\tt arXiv:hep-th/9711200}];
E.~Witten,
Adv.\ Theor.\ Math.\ Phys.\  {\bf 2}, 253 (1998)
[{\tt arXiv:hep-th/9802150}];
S.~S.~Gubser, I.~R.~Klebanov and A.~M.~Polyakov,
Phys.\ Lett.\ B {\bf 428}, 105 (1998)
[{\tt arXiv:hep-th/9802109}].


\bibitem{gu}
See for example, S.~S.~Gubser,
Phys.\ Rev.\ D {\bf 63}, 084017 (2001)
[{\tt arXiv:hep-th/9912001}].

\bibitem{apr}
N.~Arkani-Hamed, M.~Porrati and L.~Randall,
JHEP {\bf 0108}, 017 (2001)
[{\tt arXiv:hep-th/0012148}].

\bibitem{rzp}
R.~Rattazzi and A.~Zaffaroni,
JHEP {\bf 0104}, 021 (2001)
[{\tt arXiv:hep-th/0012248}];
M.~Perez-Victoria,
JHEP {\bf 0105}, 064 (2001)
[{\tt arXiv:hep-th/0105048}].


\bibitem{largen}
E.~Witten,
Nucl.\ Phys.\ B {\bf 160}, 57 (1979).

\bibitem{ls}
M.~A.~Luty and R.~Sundrum,
Phys.\ Rev.\ D {\bf 65}, 066004 (2002)
[{\tt arXiv:hep-th/0105137}];
{\tt arXiv:hep-th/0111231}.

\bibitem{pomarol}
A.~Pomarol,
Phys.\ Rev.\ Lett.\  {\bf 85}, 4004 (2000)
[{\tt arXiv:hep-ph/0005293}].

\bibitem{gp1}
T.~Gherghetta and A.~Pomarol,
Nucl.\ Phys.\ B {\bf 586}, 141 (2000)
[{\tt arXiv:hep-ph/0003129}].

\bibitem{gp2}
T.~Gherghetta and A.~Pomarol,
Nucl.\ Phys.\ B {\bf 602}, 3 (2001)
[{\tt arXiv:hep-ph/0012378}].

\bibitem{mp}
D.~Marti and A.~Pomarol,
Phys.\ Rev.\ D {\bf 64}, 105025 (2001)
[{\tt arXiv:hep-th/0106256}].

\bibitem{WarpedMSSM}
J.~A.~Casas, J.~R.~Espinosa and I.~Navarro,
Nucl.\ Phys.\ B {\bf 620}, 195 (2002)
[{\tt arXiv:hep-ph/0109127}];
W.~D.~Goldberger, Y.~Nomura and D.~R.~Smith,
{\tt arXiv:hep-ph/0209158};
K.~W.~Choi, D.~Y.~Kim, I.~W.~Kim and T.~Kobayashi,
{\tt arXiv:hep-ph/0301131};
Z.~Chacko and E.~Ponton,
{\tt arXiv:hep-ph/0301171}.

\bibitem{fayet}
P.~Fayet,
Phys.\ Lett.\ B {\bf 64}, 159 (1976).

\bibitem{gk}
A.~K.~Grant and Z.~Kakushadze,
Phys.\ Lett.\ B {\bf 465}, 108 (1999)
[{\tt arXiv:hep-ph/9906556}].

\bibitem{mapo}
M.~Masip and A.~Pomarol,
Phys.\ Rev.\ D {\bf 60}, 096005 (1999)
[{\tt arXiv:hep-ph/9902467}].

\bibitem{rsew}
C.~Csaki, J.~Erlich and J.~Terning,
Phys.\ Rev.\ D {\bf 66}, 064021 (2002)
[{\tt arXiv:hep-ph/0203034}];
G.~Burdman,
Phys.\ Rev.\ D {\bf 66}, 076003 (2002)
[{\tt arXiv:hep-ph/0205329}];
H.~Davoudiasl, J.~L.~Hewett and T.~G.~Rizzo,
{\tt arXiv:hep-ph/0212279};
M.~Carena, E.~Ponton, T.~M.~Tait and C.~E.~Wagner,
{\tt arXiv:hep-ph/0212307}.

\bibitem{dpq}
A.~Delgado, A.~Pomarol and M.~Quiros,
JHEP {\bf 0001}, 030 (2000)
[{\tt arXiv:hep-ph/9911252}].

\bibitem{gw}
W.~D.~Goldberger and M.~B.~Wise,
Phys.\ Rev.\ Lett.\  {\bf 83}, 4922 (1999)
[{\tt arXiv:hep-ph/9907447}].

\bibitem{rsm}
M.~A.~Luty and R.~Sundrum,
Phys.\ Rev.\ D {\bf 64}, 065012 (2001)
[{\tt arXiv:hep-th/0012158}];
A.~Falkowski, Z.~Lalak and S.~Pokorski,
Nucl.\ Phys.\ B {\bf 613}, 189 (2001)
[{\tt arXiv:hep-th/0102145}].

\bibitem{gapo}
J.~Garriga and A.~Pomarol,
{\tt arXiv:hep-th/0212227}.

\bibitem{higgsino}
ALEPH, DELPHI, L3 and OPAL experiments, note LEPSUSYWG/02-04.1
(http://lepsusy.web.cern.ch/lepsusy/Welcome.html). 


\end{thebibliography}
\end{document}